\PassOptionsToPackage{table,dvipsnames}{xcolor}
\documentclass[conference]{IEEEtran}
\IEEEoverridecommandlockouts

\newtheorem{definition}{Definition}
\newtheorem{theorem}{Theorem}
\newtheorem{lemma}{Lemma}

\usepackage[left=0.67in, right=0.68in,top=0.73 in,bottom = 1.05 in]{geometry}
\usepackage{cite}
\usepackage{amsmath,amssymb,amsfonts}
\usepackage{algorithmic}
\usepackage{graphicx}
\usepackage{textcomp}
\usepackage{xcolor}
\usepackage{enumitem}
\usepackage{mathtools}
\PassOptionsToPackage{hyphens}{url}\usepackage[hidelinks]{hyperref}
\usepackage{tikz,pgfplots}
\usepackage{grffile}
\pgfplotsset{compat=newest}
\usepgfplotslibrary{patchplots}
\usetikzlibrary{backgrounds,plotmarks,fit,positioning,arrows,shapes,shapes.multipart,calc,arrows.meta}
\usetikzlibrary{fit,positioning,arrows,shapes,shapes.multipart,calc,arrows.meta,shapes.geometric,shapes.misc}
\usetikzlibrary{decorations.pathreplacing,angles,quotes,calligraphy}
\usetikzlibrary{automata}
\newlength{\radx}
\newlength{\rady}

\allowdisplaybreaks

\usepackage{subfigure}

\usepackage[toc,acronym,nonumberlist,nopostdot]{glossaries}
\makeindex
\makenoidxglossaries
\glsdisablehyper
\newacronym{KL}{KL}{Kullback-Leibler}
\newacronym{DP}{DP}{differential privacy}
\newacronym{NDP}{NDP}{noiseless differential privacy}
\newacronym{PLRV}{PLRV}{privacy-loss random variable}
\newacronym{SPA}{SPA}{saddle-point accountant}
\newacronym{FL}{FL}{federated learning}

\newacronym{SecAgg}{SecAgg}{secure aggregation}
\newacronym{FP}{FP}{false positive}
\newacronym{FN}{FN}{false negative}
\newacronym{FPR}{FPR}{false positive rate}
\newacronym{FNR}{FNR}{false negative rate}

\newacronym{PDF}{PDF}{probability density function}
\newacronym{PMF}{PMF}{probability mass function}
\newacronym{CDF}{CDF}{cummulative distribution function}
\newacronym{CGF}{CGF}{cummulant generating function}

\newacronym{iid}{i.i.d.}{independent and identically distributed}

\newacronym{LLR}{LLR}{log-likelihood ratio}

\newacronym{LDP}{LDP}{local DP}
\newacronym{LIP}{LIP}{local information privacy}
\newacronym{MIDP}{MIDP}{mutual-information DP}
\newacronym{MIP}{MIP}{mutual-information privacy}
\newacronym{LMIDP}{LMIDP}{local MIDP}
\newacronym{LMIP}{LMIP}{local MIP}

\newacronym{CD}{CD}{context-dependent}
\newacronym{CI}{CI}{context-independent}

\newacronym{AWGN}{AWGN}{additive white Gaussian noise}

\usepackage{vmr-symbols-vecbold}
\usepackage{standard-macros}

\renewcommand{\triangleq}{=}
\newcommand{\KL}{\opK\opL}
\newcommand{\eps}{\epsilon}

\newcommand{\deltaOptLIP}{\bar{\delta}\supp{LIP}}
\newcommand{\deltaOptLDP}{\bar{\delta}\supp{LDP}}
\newcommand{\deltaLIP}{\delta\supp{LIP}}
\newcommand{\deltaLDP}{\delta\supp{LDP}}
\newcommand{\muCI}{\mu\supp{CI}}
\newcommand{\muCD}{\mu\supp{CD}}
\newcommand{\muOptCI}{\bar{\mu}\supp{CI}}
\newcommand{\muOptCD}{\bar{\mu}\supp{CD}}

\newcommand{\Ber}[1]{{\rm Ber}\lefto(#1\right)}

\newcommand{\Exp}[2][]{\mathbb{E}_{#1}\left[#2\right]}
\renewcommand{\Prob}[2][]{\mathbb{P}_{#1}\left[#2\right]}

\newcommand{\revise}[1]{{#1}}

\def\BibTeX{{\rm B\kern-.05em{\sc i\kern-.025em b}\kern-.08em
    T\kern-.1667em\lower.7ex\hbox{E}\kern-.125emX}}
    
\begin{document}

\title{\revise{On Local Mutual-Information Privacy}  
	\thanks{
 \revise{This work was partially supported by the 
Wallenberg AI, Autonomous Systems and Software Program
(WASP) and by the Swedish Research Council (VR) under grant 2023-05065.}}
}

\author{
\IEEEauthorblockN{Khac-Hoang Ngo\IEEEauthorrefmark{1}, Johan \"Ostman\IEEEauthorrefmark{2}, and Alexandre Graell i Amat\IEEEauthorrefmark{1}}
 \IEEEauthorblockA{\IEEEauthorrefmark{1}Department of Electrical Engineering, Chalmers University of Technology, 41296 Gothenburg, Sweden}
 \IEEEauthorblockA{\IEEEauthorrefmark{2}AI Sweden, 41756 Gothenburg, Sweden} \vspace{-.6cm}
}

\maketitle

\begin{abstract}
    \revise{Local mutual-information privacy (LMIP)} is a privacy notion that aims to quantify the reduction of uncertainty about the input data when the output of a privacy-preserving mechanism  is revealed. We study the relation of \revise{LMIP} with local differential privacy (LDP)\textemdash the \emph{de facto} standard notion of privacy in  context-independent scenarios\textemdash, and with local information privacy (LIP)\textemdash the state-of-the-art notion   for  context-dependent settings. We establish explicit conversion rules, i.e., bounds on the privacy parameters for a \revise{LMIP} mechanism to also satisfy LDP/LIP, and vice versa. 
    We use our bounds to formally verify that \revise{LMIP} is a weak privacy notion. 
    We also show that uncorrelated Gaussian noise is the best-case noise in terms of context-independent 
 \revise{LMIP} 
    if both the input data and the noise are subject to an average power constraint. 
\end{abstract}



\section{Introduction} \label{sec:intro}

Modern data-driven services heavily rely on the utilization of data distributed across various clients. This data serves various purposes, such as collaborative training of machine learning models as in \gls{FL}~\cite{McM17}, or performing federated analytics~\cite{Elk23}, under the orchestration of a central server. However, the data sharing between the clients and the central server poses severe privacy risks. For example, in \gls{FL}, a curious server may infer sensitive information about the clients from the local updates~\cite{Gei20,Nas19}. To mitigate such risks, it is imperative for the clients to apply local privacy-preserving mechanisms before sharing their data.

Various privacy notions have been proposed to characterize the privacy guarantees of such mechanisms. \Gls{DP} is a rigorous privacy measure that quantifies the ability of an adversary to guess which dataset, out of two neighboring ones, a model was trained on~\cite{Dwo06_calibrating,Dwo14}. This is typically achieved by adding noise to the model/gradients obtained from the dataset~\cite{Aba16}. For the local setting, a variant of \gls{DP} is \gls{LDP}~\cite{Kas11,Duc13}, where the noise is added to individual data points. When applied to 
\gls{FL}, \gls{LDP} lets the clients add noise to their updates before sending them to the server. \gls{LDP} is \gls{CI}, i.e., it is oblivious to the underlying data distribution. A recently proposed \gls{CD} local privacy notion is \gls{LIP}~\cite{Bo21}, which guarantees that the ratio between the posterior and prior of the input data is bounded. 

Information leakage can also be captured by the mutual information between the input data and the output of the privacy-preserving mechanism. This gives rise to the \revise{\gls{MIP}} notion~\cite{Cuf16},\footnote{\revise{This notion is called mutual-information \gls{DP} in~\cite{Cuf16}. Here, we drop the term ``differential'' since the concept of differentiating two neighboring datasets is not explicit in the definition.}} which can be directly adapted to the local setting by regarding the input as the local data. \revise{\Gls{LMIP}} can be either \gls{CD} or \gls{CI}, depending on whether the bound on the information leakage is applied to the mutual information for a given data distribution or for all possible data distributions. The \revise{authors of}~\cite{Cuf16} analyze the relation between \gls{CI}-\revise{\gls{MIP}} and \gls{DP}, but \revise{rely} on a crude ordering of privacy metrics that does not lead to explicit conversion rules. In~\cite{Wan16}, the relation between \gls{CD}-\revise{\gls{MIP}} and \gls{DP} is analyzed, while~\cite{Bo21,Bo22} show that \gls{LIP} implies \gls{CD}-\revise{\gls{LMIP}}. However, these works are restricted to a special setting \revise{where, implicitly, the failure probability of the privacy constraint, represented by a parameter~$\delta$, is $0$. Note that allowing for $\delta > 0$ enables the design of important mechanisms, such as the Gaussian mechanism~\cite{Dwo06}.} 

In this paper, we put forth explicit relations between \gls{CI}-\revise{\gls{LMIP}} and \gls{LDP}, and between \gls{CD}-\revise{\gls{LMIP}} and \gls{LIP} \revise{in the case with $\delta > 0$}. We provide closed-form conversion rules, i.e., bounds on the privacy parameters for a \revise{\gls{CI}/\gls{CD}-\gls{LMIP}} mechanism to also satisfy \gls{LDP}/\gls{LIP}, and vice versa. Using our bounds, we formalize and put in plain sight the existing claim that \revise{\gls{LMIP}} is a weak privacy notion~\cite{Bo21,Bo22}. Specifically, we show that a mechanism designed to achieve a low mutual-information leakage need not achieve a strong \gls{LDP}/\gls{LIP} guarantee. Therefore, \revise{\gls{LMIP}} should not be used as an objective when designing privacy-preserving mechanisms. However, the mutual-information leakage can still be used as a tool to gain insights into the properties of a mechanism, as done in~\cite{Cuf16}. Notably, through the lens of \gls{CI}-\revise{\gls{LMIP}}, we show that uncorrelated Gaussian noise is the best-case noise for an output perturbation mechanism if both the input data and the noise are subject to an average power constraint. This result supports the widespread use of the Gaussian mechanism.

\subsubsection*{Notation}
Uppercase italic letters, e.g.,~$X$, denote scalar random variables and their realizations are written in lowercase, e.g.,~$x$. Vectors are denoted likewise with boldface letters, e.g., a random vector~$\rvecx$ and its realization~$\vecx$. Uppercase sans-serif letters, e.g., $\matX$, denote deterministic matrices. We denote the $d \times d$ identity matrix by~$\matidentity_d$. Calligraphic letters denote sets or events. 
We denote the Bernoulli distribution with parameter $p$ by $\Ber{p}$. 
Furthermore, $\ind{\cdot}$ denotes the indicator function and $\log$ the base-$2$ logarithm.
We define the hockey-stick divergence with parameter $\alpha$ between two probability measures $P$ and $Q$ as
        $\opH_\alpha(P\|Q) = 
         \sup_{\setE} (P(\setE) - \alpha Q(\setE))$. 

\section{Local Privacy Notions} \label{sec:notions}





Consider a distributed system where each \revise{client} applies a privacy mechanism $M(\cdot)$ to its data $\rvecx$ and shares $M(\rvecx)$ with an untrusted server. A local privacy notion is used to characterize the privacy guarantee of such mechanism. The focus of our work is on \revise{\gls{LMIP}} and its relation with other local privacy notions. 
We consider two versions of \revise{\gls{LMIP}}, namely, \gls{CI}-\revise{\gls{LMIP}} and \gls{CD}-\revise{\gls{LMIP}}.
\begin{definition}[\gls{CI}-\revise{\gls{LMIP}}]  \label{def:CI-LMIDP} 
    For a data distribution space $\setP$, a randomized mechanism $M$ satisfies $\mu$-\gls{CI}-\revise{\gls{LMIP}} if and only if 
    \begin{equation}
        \sup_{P_{\rvecx} \in \setP} I(\rvecx; M(\rvecx)) \le \mu~~\text{bits}.
    \end{equation}
\end{definition}
\begin{definition}[\gls{CD}-\revise{\gls{LMIP}}]  \label{def:CD-LMIDP} 
    For a data distribution $P_\rvecx$, a randomized mechanism $M$ satisfies $\mu$-\gls{CD}-\revise{\gls{LMIP}} if and only if 
    \begin{equation}
        I(\rvecx; M(\rvecx)) \le \mu~~\text{bits}.
    \end{equation}
\end{definition}

\revise{\gls{LMIP}} aims to quantify the reduction of uncertainty about $\rvecx$ when $M(\rvecx)$ is revealed. \revise{We obtain \gls{CI}-\revise{\gls{LMIP}} by adapting \revise{\gls{MIP}}, which was proposed for the centralized setting~\cite[Def.~2]{Cuf16}, to the local setting. \gls{CI}-\gls{LMIP}} aims to characterize the privacy for all possible data distributions, while \gls{CD}-\revise{\gls{LMIP}} addresses a fixed data distribution. 


For the \gls{CI} setting, the \emph{de facto} standard notion of local privacy  is \gls{LDP}, defined as follows. 
\begin{definition}[{\Gls{LDP}~\cite{Kas11,Duc13}}] \label{def:LDP}
    A randomized mechanism $M$ satisfies $(\epsilon,\delta)$-\gls{LDP} if and only if, for every pair of data points $(\vecx,\vecx')$ and for every measurable set $\setE$, we have that
    \begin{equation}
        \Prob{M(\vecx) \in \setE} \le e^{\epsilon} \Prob{M(\vecx') \in \setE} + \delta, \label{eq:def_LDP}
    \end{equation}
    or, equivalently, 
    \begin{equation}
        \sup_{\vecx \ne \vecx'} \opH_{e^\epsilon}(P_{M(\vecx)} \| P_{M(\vecx')}) \le \delta. \label{eq:LDP_hockeystick}
    \end{equation}
\end{definition}

\Gls{LDP} aims to characterize the ability of a server that observes the mechanism output to find which input, out of two possible ones, was used by the client.

For the \gls{CD} setting, a recently proposed notion of local privacy is \gls{LIP}, defined as follows.
\begin{definition}[\Gls{LIP}~\cite{Bo21}] \label{def:LIP}
    For a data distribution $P_\rvecx$, a randomized mechanism $M$ satisfies $(\epsilon,\delta)$-\gls{LIP} if and only if, for every data point $\vecx$ and every measurable set $\setE$, we have that
    \begin{multline} 
        e^{-\epsilon} \Prob{M(\vecx) \in \setE} - \delta \le \Prob{M(\rvecx) \in \setE}  \\ 
        \le e^\epsilon \Prob{M(\vecx) \in \setE} + \delta, \label{eq:def_LIP}
    \end{multline}
    or, equivalently, 
    \begin{multline}
        \sup_{\vecx} \max\big\{\opH_{e^\epsilon}(P_{M(\rvecx)} \| P_{M(\vecx)}), e^{-\epsilon} 
         \opH_{e^\epsilon}(P_{M(\vecx)} \| P_{M(\rvecx)}) \big\}  \\ \le \delta.  \label{eq:LIP_hockeystick}
    \end{multline}
\end{definition}

\Gls{LIP} aims to guarantee that the mechanism output provides limited information about any possible input $\vecx$. 

For a mechanism $M$, we define the optimal \gls{LDP} curve $\deltaOptLDP_M(\eps)$ and the optimal \gls{LIP} curve $\deltaOptLIP_M(\eps)$ as the functions that return the smallest value of $\delta$ for which $M$ satisfies $(\eps,\delta)$-\gls{LDP} and $(\eps,\delta)$-\gls{LIP}, respectively, \revise{for a given $\epsilon$}. Furthermore, we define the optimal \gls{CI}-\revise{\gls{LMIP}} parameter $\muOptCI_M$ and the optimal \gls{CD}-\revise{\gls{LMIP}} parameter $\muOptCD_M$ as the smallest values of $\mu$ such that $M$ satisfies $\mu$-\gls{CI}-\revise{\gls{LMIP}} and $\mu$-\gls{CD}-\revise{\gls{LMIP}}, respectively. 

    Throughout the paper, we will repeatedly consider the Gaussian mechanism
    \begin{equation}
        G(\rvecx) = \rvecx +  \normal(\mathbf{0}, \sigma^2\matidentity_d). \label{eq:Gaussian_mechanism}
    \end{equation}
    We assume that $\rvecx \in \reals^d$ and $\|\rvecx\| \le \sqrt{d}\Delta$. The optimal \gls{CI}-\revise{\gls{LMIP}} parameter of $G$ is 
    \begin{equation}
        \muOptCI_G = \sup_{P_{\rvecx}\colon \|\rvecx\| \le \sqrt{d}\Delta} I(\rvecx; \rvecx + \normal(\mathbf{0}, \sigma^2\matidentity_d)),
    \end{equation} 
    which \revise{is the capacity of a vector \gls{AWGN} channel with peak power constraint} and can be computed based on~\cite{Ras16}. Furthermore, it follows from~\cite{Bal18} that the optimal \gls{LDP} curve of $G$ is 
    \begin{equation}
        \deltaOptLDP_G(\epsilon) = \Phi\bigg(\frac{\sqrt{d}\Delta}{\sigma}-\frac{\epsilon \sigma}{2\sqrt{d}\Delta}\bigg) - e^{\epsilon} \Phi\bigg(- \frac{\sqrt{d}\Delta}{\sigma}-\frac{\epsilon \sigma}{2\sqrt{d}\Delta}\bigg),
    \end{equation} 
    where $\Phi(x) = \frac{1}{\sqrt{2\pi}} \int_{-\infty}^x e^{-u^2/2} \dif u$ is the cummulative distribution function of the standard normal distribution. 

    Now consider the case $d = 1$, i.e., $G(X) = X + \normal(0,\sigma^2)$, and assume further that the input $X$ follows a discrete distribution in $\{x_1,\dots,x_n\} \subset [-\Delta,\Delta]$. We have that $G(x_i) \sim \normal(x_i, \sigma^2)$ and $G(X)$ follows the Gaussian mixture model $\sum_{i=1}^n P_X(x_i) \normal(x_i, \sigma^2)$. The optimal \gls{CD}-\revise{\gls{LMIP}} parameter $\muOptCD_G$ and optimal \gls{LIP} curve $\deltaOptLIP_G(\epsilon)$ of $G$ can be computed based on numerical evaluation of the \gls{KL} and hockey-stick divergences between $P_{G(X)}$ and $P_{G(x)}$. 

\section{Relation Between \revise{\gls{LMIP}}, \gls{LDP}, and \gls{LIP}} \label{sec:relation}
In this section, we report some explicit relations between the local privacy notions mentioned in Section~\ref{sec:notions}. 

\subsection{\gls{CI}-\revise{\gls{LMIP}} and \gls{LDP}}
We first consider the \gls{CI} notions, i.e., \gls{CI}-\revise{\gls{LMIP}} and \gls{LDP}.
\begin{theorem}[\gls{CI}-\revise{\gls{LMIP}} vs. \gls{LDP}] \label{th:LMIDP_and_LDP}
    \begin{enumerate}[label=(\alph*),leftmargin=*]
        \item \label{th:LMIDP_implies_LDP} If a mechanism $M$ satisfies $\mu$-\gls{CI}-\revise{\gls{LMIP}}, it satisfies $(\epsilon,\deltaLDP_\mu(\epsilon))$-\gls{LDP} for every $\epsilon > 0$ and 
        \begin{align}
        \deltaLDP_\mu(\epsilon) = &\max_{p_0,p_1 \in [0,1]} \max\{0, p_0 - e^\epsilon p_1, p_1 - e^\epsilon p_0\}, \label{eq:delta_mu}\\
            &\textnormal{\revise{subject to}}~~ C\sub{BAC}(p_0,1-p_1) \le \mu \notag
        \end{align}
        where $C\sub{BAC}(\epsilon_0,\epsilon_1)$ is the capacity of the binary asymmetric channel with crossover probabilities $(\epsilon_0,\epsilon_1)$~\cite{Mos09}. Specifically, 
        if $0 \le \epsilon_0 \le \min\{\epsilon_1, 1 - \epsilon_1, 1/2\}$,
        \begin{multline}
            C\sub{BAC}(\epsilon_0,\epsilon_1) =
                \tfrac{\epsilon_0}{1 - \epsilon_0 - \epsilon_1} H\sub{b}(\epsilon_1) - \tfrac{1 - \epsilon_1}{1 - \epsilon_0-\epsilon_1} H\sub{b}(\epsilon_0) \\ + \log\big(1 + 2^{ \frac{H\sub{b}(\epsilon_0) - H\sub{b}(\epsilon_1)}{1 - \epsilon_0 - \epsilon_1}}\big),
        \end{multline}
        where $H\sub{b}(p) \triangleq -p\log p - (1\!-p)\log (1-p)$  is the binary entropy.
        Otherwise, $C\sub{BAC}(\epsilon_0,\epsilon_1)$ is equal to $C\sub{BAC}(\epsilon_1,\epsilon_0)$ if $\epsilon_0 > \epsilon_1$, to $C\sub{BAC}(1 - \epsilon_1,1 - \epsilon_0)$ if $\epsilon_0 > 1 - \epsilon_1$, and to $C\sub{BAC}(1 - \epsilon_0,1 - \epsilon_1)$ if $\epsilon_0 > 1/2$. 
        A mechanism that satisfies $\mu$-\gls{CI}-\revise{\gls{LMIP}} need not satisfy $(\epsilon,\delta)$-\gls{LDP} for $\epsilon > 0$ and any $\delta < \deltaLDP_{\mu}(\epsilon)$.

        \item \label{th:LDP_implies_LMIDP} If a mechanism $M$ satisfies $(\epsilon,\delta(\epsilon))$-\gls{LDP} for $\epsilon > 0$, it satisfies $\muCI_{\delta(\epsilon)}$-\gls{CI}-\revise{\gls{LMIP}} with 
        \begin{equation} \label{eq:mu_delta}
            \muCI_{\delta(\epsilon)} = \log(e) \int_0^\infty (1 + e^{-\epsilon}) \delta(\epsilon) \dif \epsilon \quad \text{bits}.
        \end{equation}
    \end{enumerate}
\end{theorem}

\begin{IEEEproof}
    See Appendix~\ref{proof:LMIDP_and_LDP}.
\end{IEEEproof}

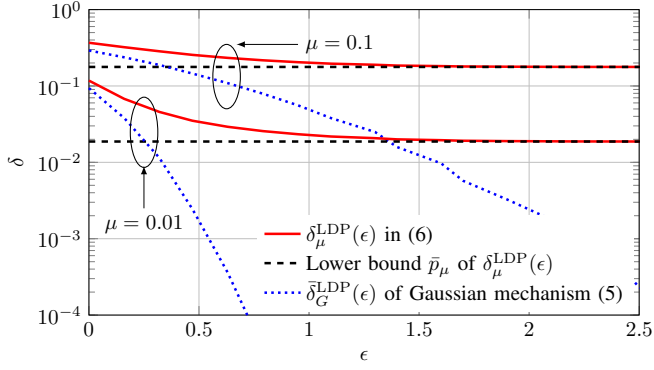
\begin{figure}[t]
    \centering
\begin{tikzpicture}[scale = .8]
    \begin{axis}[%
        width=3.6in,
        height=2in,
        at={(0.759in,0.481in)},
        scale only axis,
        xmin=0,
        xmax=2.5,
        xlabel style={font=\color{black},yshift=1ex},
        xlabel={$\epsilon$},
        ymin=1e-4,
        ymax=1,
        ymode=log,
        yminorticks=true,
        ylabel={$\delta$},
        xmajorgrids,
        ymajorgrids,
        legend style={at={(.99,.01)}, anchor=south east, legend cell align=left, align=left,draw=none, fill=white, fill opacity=.3,text opacity = 1}
        ]		





        
        \addplot [color=red,line width=1.2pt]
        table[row sep=crcr]{%
            2.2e-16 3.6796e-01 \\ 
1.6e-01 3.2200e-01 \\ 
3.2e-01 2.8457e-01 \\ 
4.7e-01 2.5542e-01 \\ 
6.3e-01 2.3335e-01 \\ 
7.9e-01 2.1689e-01 \\ 
9.5e-01 2.0476e-01 \\ 
1.1e+00 1.9590e-01 \\ 
1.3e+00 1.8953e-01 \\ 
1.4e+00 1.8505e-01 \\ 
1.6e+00 1.8200e-01 \\ 
1.7e+00 1.8001e-01 \\ 
1.9e+00 1.7879e-01 \\ 
2.1e+00 1.7810e-01 \\ 
2.2e+00 1.7775e-01 \\ 
2.4e+00 1.7759e-01 \\ 
2.5e+00 1.7753e-01 \\ 
2.7e+00 1.7751e-01 \\ 
2.8e+00 1.7750e-01 \\ 
3.0e+00 1.7750e-01 \\ 
            };
        \addlegendentry{$\deltaLDP_\mu(\epsilon)$ in~\eqref{eq:delta_mu}};

        \addplot [color=black,dashed,line width=1.2pt]
        table[row sep=crcr]{%
            2.2e-16 1.7750e-01 \\ 
            5 1.7750e-01 \\ 
            };
        \addlegendentry{Lower bound $\bar{p}_\mu$ of $\deltaLDP_\mu(\epsilon)$};

        \addplot [color=blue,dotted,line width=1.2pt]
        table[row sep=crcr]{%
            2.2e-16 2.9132e-01 \\ 
1.6e-01 2.3740e-01 \\ 
3.2e-01 1.8844e-01 \\ 
4.7e-01 1.4545e-01 \\ 
6.3e-01 1.0901e-01 \\ 
7.9e-01 7.9222e-02 \\ 
9.5e-01 5.5753e-02 \\ 
1.1e+00 3.7954e-02 \\ 
1.3e+00 2.4966e-02 \\ 
1.4e+00 1.5855e-02 \\ 
1.6e+00 9.7121e-03 \\ 
1.7e+00 5.7343e-03 \\ 
1.9e+00 3.2612e-03 \\ 
2.1e+00 1.7854e-03 \\ 
2.2e+00 9.4040e-04 \\ 
2.4e+00 4.7633e-04 \\ 
2.5e+00 2.3192e-04 \\ 
2.7e+00 1.0850e-04 \\ 
2.8e+00 4.8752e-05 \\ 
3.0e+00 2.1034e-05 \\ 
            };
        \addlegendentry{$\deltaOptLDP_G(\epsilon)$ of Gaussian mechanism~\eqref{eq:Gaussian_mechanism}};

        \addplot [color=red,line width=1.2pt,forget plot]
        table[row sep=crcr]{%
            2.2e-16 1.1760e-01 \\ 
1.6e-01 6.7883e-02 \\ 
3.2e-01 4.5727e-02 \\ 
4.7e-01 3.5079e-02 \\ 
6.3e-01 2.9212e-02 \\ 
7.9e-01 2.5645e-02 \\ 
9.5e-01 2.3335e-02 \\ 
1.1e+00 2.1784e-02 \\ 
1.3e+00 2.0723e-02 \\ 
1.4e+00 1.9996e-02 \\ 
1.6e+00 1.9504e-02 \\ 
1.7e+00 1.9179e-02 \\ 
1.9e+00 1.8974e-02 \\ 
2.1e+00 1.8852e-02 \\ 
2.2e+00 1.8785e-02 \\ 
2.4e+00 1.8751e-02 \\ 
2.5e+00 1.8736e-02 \\ 
2.7e+00 1.8732e-02 \\ 
2.8e+00 1.8730e-02 \\ 
3.0e+00 1.8730e-02 \\ 
            };

        \addplot [color=black,dashed,line width=1.2pt,forget plot]
        table[row sep=crcr]{%
            2.2e-16 1.8730e-02 \\ 
            5 1.8730e-02 \\ 
            };

        \addplot [color=blue,dotted,line width=1.2pt,forget plot]
        table[row sep=crcr]{%
            2.2e-16 9.3759e-02 \\ 
1.6e-01 3.8143e-02 \\ 
3.2e-01 1.1458e-02 \\ 
4.7e-01 2.4484e-03 \\ 
6.3e-01 3.6206e-04 \\ 
7.9e-01 3.6329e-05 \\ 
9.5e-01 2.4390e-06 \\ 
1.1e+00 1.0846e-07 \\ 
1.3e+00 3.1706e-09 \\ 
1.4e+00 6.0601e-11 \\ 
1.6e+00 7.5409e-13 \\ 
1.7e+00 6.0894e-15 \\ 
1.9e+00 3.1829e-17 \\ 
2.1e+00 1.0747e-19 \\ 
2.2e+00 2.3401e-22 \\ 
2.4e+00 3.2819e-25 \\ 
2.5e+00 2.9610e-28 \\ 
2.7e+00 1.7171e-31 \\ 
2.8e+00 6.3952e-35 \\ 
3.0e+00 1.5287e-38 \\ 
            };

            
            
        \coordinate (Center) at (axis cs:.42,2.3e-2);
   		\coordinate (Radius) at (axis cs:.47,5e-2);
   		
   		\coordinate (Center2) at (axis cs:.12,6e-3);
   		\coordinate (Radius2) at (axis cs:.17,1.4e-2);

        \draw[-latex] (axis cs:.95,3.5e-1) node[right] {$\mu = 0.1$} -- (axis cs:.67,3.5e-1);
        
        \draw[-latex] (axis cs:.25,2.5e-3) node[below] {$\mu = 0.01$} -- (axis cs:.25,9e-3);
        
    \end{axis}
    \pgfextractx{\radx}{\pgfpointdiff{\pgfpointanchor{Radius}{center}}{\pgfpointanchor{Center}{center}}}%
	\pgfextracty{\rady}{\pgfpointdiff{\pgfpointanchor{Radius}{center}}{\pgfpointanchor{Center}{center}}}%
	\draw (Center) ellipse [x radius = \radx, y radius = \rady]; 
	
	\pgfextractx{\radx}{\pgfpointdiff{\pgfpointanchor{Radius2}{center}}{\pgfpointanchor{Center2}{center}}}%
	\pgfextracty{\rady}{\pgfpointdiff{\pgfpointanchor{Radius2}{center}}{\pgfpointanchor{Center2}{center}}}%
	\draw (Center2) ellipse [x radius = \radx, y radius = \rady]; 
\end{tikzpicture}%
\vspace{-1cm}


    \caption{The achievable \gls{LDP} curve $\deltaLDP_\mu(\epsilon)$ of a $\mu$-\gls{CI}-\revise{\gls{LMIP}} mechanism, its lower bound $\bar{p}_\mu$, and the optimal \gls{LDP} curve $\deltaOptLDP_G(\epsilon)$ of the Gaussian mechanism $G$ in~\eqref{eq:Gaussian_mechanism} with $d = 10$, $\Delta = 1$, and $\sigma^2$ chosen such that $\muOptCI_G = \mu$.}
    \label{fig:LMIDP_LDP}
    \vspace{-.3cm}
\end{figure}

Theorem~\ref{th:LMIDP_and_LDP}\ref{th:LMIDP_implies_LDP} implies that $\deltaLDP_\mu(\epsilon)$ is the smallest value of~$\delta$ such that every mechanism $M$ satisfying $\mu$-\gls{CI}-\revise{\gls{LMIP}} also satisfies $(\epsilon,\delta)$-\gls{LDP}. We next provide some remarks and demonstrate them in Fig~\ref{fig:LMIDP_LDP}.

\begin{itemize} 
    \item For a fixed $\epsilon$, a smaller $\mu$ leads to a smaller $\deltaLDP_\mu(\epsilon)$. Indeed, a stronger mechanism in terms of \revise{\gls{LMIP}} also provides a stronger \gls{LDP} guarantee.  
    
    \item If $\mu \ge 1$ bit, then $\deltaLDP_\mu(\epsilon) = 1$, i.e., no \gls{LDP} is guaranteed. Indeed, revealing the binary value of $\ind{\rvecx \in \setA}$ for some set~$\setA$ is enough to distinguish any pair $\vecx \in \setA$ and $\vecx' \notin \setA$.

    \item If $\mu < 1$ bit, $\deltaLDP_\mu(\epsilon)$ is lower-bounded by the unique value of $\bar{p}_\mu \in [0,1]$ such that $H\sub{b}(\bar{p}_\mu)/\bar{p}_\mu = -\log(2^\mu - 1)$, and converges to this value when $\epsilon \!\to\!\infty$, as depicted in Fig.~\ref{fig:LMIDP_LDP}. That is, $\deltaLDP_\mu(\delta)$ does not vanish as $\epsilon$ becomes large. This means that \gls{CI}-\revise{\gls{LMIP}} implies only a weak \gls{LDP} guarantee.
\end{itemize}


In Fig.~\ref{fig:LMIDP_LDP}, we also plot the optimal \gls{LDP} curve $\deltaOptLDP_G(\epsilon)$ of the Gaussian mechanism $G$ in~\eqref{eq:Gaussian_mechanism} with the noise calibrated such that $\muOptCI_G = \mu$. We observe that this mechanism achieves a much lower \gls{LDP} curve than the general guarantee $\deltaLDP_\mu(\epsilon)$. Therefore, a strong \gls{CI}-\revise{\gls{LMIP}} mechanism can also have a strong \gls{LDP} guarantee. However, this is not ensured if the mechanism is designed for \gls{CI}-\revise{\gls{LMIP}} only.  

In Fig.~\ref{fig:LDP_LMIDP}, we plot $\muCI_{\delta(\epsilon)}$ for $\delta(\epsilon) = \deltaOptLDP_G(\epsilon)$ with $\Delta = 1$, $d \in \{1,10\}$, and $\sigma^2 \in [0,200]$. We also show the optimal \gls{CI}-\revise{\gls{LMIP}} parameter $\muOptCI_G$ of the corresponding Gaussian mechanism. The gap between $\muCI_{\delta(\epsilon)}$ and $\muOptCI_G$ remains almost constant, and $\muCI_{\delta_G(\epsilon)}$ keeps decreasing as $\sigma^2$ grows. This shows that designing a strong \gls{LDP} mechanism 
also leads to a strong \gls{CI}-\revise{\gls{LMIP}} mechanism. 
\begin{figure}
    \centering
    \begin{tikzpicture}[scale = .8]
    \begin{axis}[%
        width=3.6in,
        height=2in,
        at={(0.759in,0.481in)},
        scale only axis,
        xmin=0,
        xmax=200,
        xlabel style={font=\color{black},yshift=.5ex},
        xlabel={noise variance $\sigma^2$},
        ymin=1e-3,
        ymax=1,
        yminorticks=true,
        ymode=log,
        ylabel={$\mu$},
        xmajorgrids,
        ymajorgrids,
        legend style={at={(.01,.01)}, anchor=south west, legend cell align=left, align=left,draw=none, fill=white, fill opacity=.3,text opacity = 1}
        ]		
        
        \addplot [color=red,line width=1.2pt]
        table[row sep=crcr]{%
            1.0e+00 2.8854e+00 \\ 
2.0e+00 1.4427e+00 \\ 
4.0e+00 7.2135e-01 \\ 
6.0e+00 4.8090e-01 \\ 
8.0e+00 3.6067e-01 \\ 
1.0e+01 2.8854e-01 \\ 
2.0e+01 1.4427e-01 \\ 
3.0e+01 9.6180e-02 \\ 
4.0e+01 7.2135e-02 \\ 
5.0e+01 5.7708e-02 \\ 
6.0e+01 4.8090e-02 \\ 
7.0e+01 4.1220e-02 \\ 
8.0e+01 3.6067e-02 \\ 
9.0e+01 3.2060e-02 \\ 
1.0e+02 2.8854e-02 \\ 
1.1e+02 2.6231e-02 \\ 
1.2e+02 2.4045e-02 \\ 
1.3e+02 2.2195e-02 \\ 
1.4e+02 2.0610e-02 \\ 
1.5e+02 1.9236e-02 \\ 
1.6e+02 1.8034e-02 \\ 
1.7e+02 1.6973e-02 \\ 
1.8e+02 1.6030e-02 \\ 
1.9e+02 1.5186e-02 \\ 
2.0e+02 1.4427e-02 \\ 
            };
        \addlegendentry{$\muCI_{\delta(\epsilon)}$ in~\eqref{eq:mu_delta}};

        \addplot [color=blue,dotted,line width=1.2pt]
        table[row sep=crcr]{%
            1.0e+00 5.0000e-01 \\ 
2.0e+00 2.9248e-01 \\ 
4.0e+00 1.6096e-01 \\ 
6.0e+00 1.1120e-01 \\ 
8.0e+00 8.4963e-02 \\ 
1.0e+01 6.8744e-02 \\ 
2.0e+01 3.5194e-02 \\ 
3.0e+01 2.3652e-02 \\ 
4.0e+01 1.7812e-02 \\ 
5.0e+01 1.4284e-02 \\ 
6.0e+01 1.1923e-02 \\ 
7.0e+01 1.0232e-02 \\ 
8.0e+01 8.9605e-03 \\ 
9.0e+01 7.9703e-03 \\ 
1.0e+02 7.1772e-03 \\ 
1.1e+02 6.5276e-03 \\ 
1.2e+02 5.9858e-03 \\ 
1.3e+02 5.5271e-03 \\ 
1.4e+02 5.1337e-03 \\ 
1.5e+02 4.7925e-03 \\ 
1.6e+02 4.4939e-03 \\ 
1.7e+02 4.2303e-03 \\ 
1.8e+02 3.9959e-03 \\ 
1.9e+02 3.7861e-03 \\ 
2.0e+02 3.5972e-03 \\ 
            };
        \addlegendentry{$\muOptCI_G$ of Gaussian mechanism~\eqref{eq:Gaussian_mechanism}};

        \addplot [color=red,line width=1.2pt]
        table[row sep=crcr]{%
            1.0e+00 2.8854e+01 \\ 
1.0e+01 2.8854e+00 \\ 
2.0e+01 1.4427e+00 \\ 
3.0e+01 9.6180e-01 \\ 
4.0e+01 7.2135e-01 \\ 
5.0e+01 5.7708e-01 \\ 
6.0e+01 4.8090e-01 \\ 
7.0e+01 4.1220e-01 \\ 
8.0e+01 3.6067e-01 \\ 
9.0e+01 3.2060e-01 \\ 
1.0e+02 2.8854e-01 \\ 
1.1e+02 2.6231e-01 \\ 
1.2e+02 2.4045e-01 \\ 
1.3e+02 2.2195e-01 \\ 
1.4e+02 2.0610e-01 \\ 
1.5e+02 1.9236e-01 \\ 
1.6e+02 1.8034e-01 \\ 
1.7e+02 1.6973e-01 \\ 
1.8e+02 1.6030e-01 \\ 
1.9e+02 1.5186e-01 \\ 
2.0e+02 1.4427e-01 \\  
            };

        \addplot [color=blue,dotted,line width=1.2pt]
        table[row sep=crcr]{%
            1.0e+00 4.9745e+00 \\ 
1.0e+01 6.8750e-01 \\ 
2.0e+01 3.5194e-01 \\ 
3.0e+01 2.3653e-01 \\ 
4.0e+01 1.7812e-01 \\ 
5.0e+01 1.4284e-01 \\ 
6.0e+01 1.1923e-01 \\ 
7.0e+01 1.0232e-01 \\ 
8.0e+01 8.9608e-02 \\ 
9.0e+01 7.9706e-02 \\ 
1.0e+02 7.1775e-02 \\ 
1.1e+02 6.5279e-02 \\ 
1.2e+02 5.9862e-02 \\ 
1.3e+02 5.5273e-02 \\ 
1.4e+02 5.1334e-02 \\ 
1.5e+02 4.7347e-02 \\ 
1.6e+02 4.4350e-02 \\ 
1.7e+02 4.1712e-02 \\ 
1.8e+02 3.9937e-02 \\ 
1.9e+02 3.7266e-02 \\ 
2.0e+02 3.5380e-02 \\ 
            };

   		

        \draw[-latex] (axis cs:140,1e-1) node[right] {$d = 10$} -- (axis cs:123,2.2e-1);
        \draw[-latex] (axis cs:140,1e-1) -- (axis cs:123,6.5e-2);
        
        \draw[-latex] (axis cs:140,1e-2) node[right] {$d = 1$} -- (axis cs:123,2.2e-2);
        \draw[-latex] (axis cs:140,1e-2) -- (axis cs:123,6.5e-3);
    \end{axis}
	
\end{tikzpicture}%
\vspace{-1cm}
    \caption{The achievable \gls{CI}-\revise{\gls{LMIP}} parameter $\muCI_{\delta(\epsilon)}$ of a mechanism that satisfies $(\epsilon,\delta(\epsilon))$-\gls{LDP} with $\delta(\epsilon) = \deltaOptLDP_G(\epsilon)$, and the optimal \gls{CI}-\revise{\gls{LMIP}} parameter $\muOptCI_G$ of the Gaussian mechanism~\eqref{eq:Gaussian_mechanism} with $\Delta = 1$.}
    \label{fig:LDP_LMIDP}
    \vspace{-.3cm}
\end{figure}

\subsection{\gls{CD}-\revise{\gls{LMIP}} and \gls{LIP}}
We next consider the \gls{CD} notions, i.e., \gls{CD}-\revise{\gls{LMIP}} and \gls{LIP}. 
\begin{theorem}[\gls{CD}-\revise{\gls{LMIP}} vs. \gls{LIP}] \label{th:LMIDP_and_LIP}
    \begin{enumerate}[label=(\alph*),leftmargin=*]
        \item If a mechanism $M$ satisfies $\mu$-\gls{CD}-\revise{\gls{LMIP}}, it satisfies $(\epsilon,\deltaLIP_\mu(\epsilon))$-\gls{LIP} for every $\epsilon > 0$ and 
            \begin{align}
                \deltaLIP_\mu(\epsilon) = \max_{p_0,p_1 \in [0,1]} &\max\{0, p_0 \!-\! e^\epsilon p_1, e^{-\epsilon} p_1 \!-\!  p_0\} \label{eq:delta_mu_LIP}\\
                \text{\revise{subject to}}~~ &\KL(\Ber{p_1} \| \Ber{p_0}) \le \mu. \notag
            \end{align}
        Here, {$\KL(\Ber{p_1} \| \Ber{p_0}) = (1-p_1)\log\frac{1-p_1}{1-p_0} + p_1 \log \frac{p_1}{p_0}$ is the \gls{KL} divergence between two Bernoulli distributions.}
        
        \item If a mechanism $M$ satisfies $(\epsilon,\delta(\epsilon))$-\gls{LIP} for $\epsilon > 0$, it satisfies $\muCD_{\delta(\epsilon)}$-\revise{\gls{LMIP}} with 
        \begin{equation} \label{eq:mu_delta_LIP}
                \muCD_{\delta(\epsilon)} = \log(e) \int_0^\infty (e^{\epsilon} + e^{-\epsilon}) \delta(\epsilon) \dif \epsilon \quad \text{bits}.
            \end{equation}
        \end{enumerate}
\end{theorem}
\begin{IEEEproof}
    See Appendix~\ref{proof:LMIDP_and_LIP}.
\end{IEEEproof}

\begin{figure}[t]
    \centering
\begin{tikzpicture}[scale = .8]
    \begin{axis}[%
        width=3.6in,
        height=2in,
        at={(0.759in,0.481in)},
        scale only axis,
        xmin=0,
        xmax=2.5,
        xlabel style={font=\color{black},yshift=1ex},
        xlabel={$\epsilon$},
        ymin=1e-4,
        ymax=1,
        yminorticks=true,
        ymode=log,
        ylabel={$\delta$},
        xmajorgrids,
        ymajorgrids,
        legend style={at={(1.01,.01)}, anchor=south east, legend cell align=left, align=left,draw=none, fill=white, fill opacity=.5,text opacity = 1}
        ]		





        
        \addplot [color=red,line width=1.2pt]
        table[row sep=crcr]{%
            2.2e-16 1.8544e-01 \\ 
1.6e-01 1.3748e-01 \\ 
3.2e-01 1.0890e-01 \\ 
4.7e-01 9.2302e-02 \\ 
6.3e-01 8.2384e-02 \\ 
7.9e-01 7.6278e-02 \\ 
9.5e-01 7.2463e-02 \\ 
1.1e+00 7.0088e-02 \\ 
1.3e+00 6.8645e-02 \\ 
1.4e+00 6.7805e-02 \\ 
1.6e+00 6.7348e-02 \\ 
1.7e+00 6.7121e-02 \\ 
1.9e+00 6.7021e-02 \\ 
2.1e+00 6.6983e-02 \\ 
2.2e+00 6.6971e-02 \\ 
2.4e+00 6.6968e-02 \\ 
2.5e+00 6.6967e-02 \\ 
2.7e+00 6.6967e-02 \\ 
2.8e+00 6.6967e-02 \\ 
3.0e+00 6.6967e-02 \\ 
            };
        \addlegendentry{$\deltaLIP_\mu(\epsilon)$ in~\eqref{eq:delta_mu_LIP}};

        \addplot [color=black,dashed,line width=1.2pt]
        table[row sep=crcr]{%
            0 0.0670 \\ 
            3 0.0670 \\ 
            };
        \addlegendentry{Lower bound $1-2^{-\mu}$ of $\deltaLIP_\mu(\epsilon)$};

        \addplot [color=OliveGreen,dashdotted,line width=1.2pt]
        table[row sep=crcr]{%
            2.2e-16 3.2195e-01 \\ 
1.6e-01 2.7500e-01 \\ 
3.2e-01 2.3186e-01 \\ 
4.7e-01 1.9300e-01 \\ 
6.3e-01 1.5862e-01 \\ 
7.9e-01 1.2874e-01 \\ 
9.5e-01 1.0322e-01 \\ 
1.1e+00 8.1763e-02 \\ 
1.3e+00 6.4006e-02 \\ 
1.4e+00 4.9530e-02 \\ 
1.6e+00 3.7892e-02 \\ 
1.7e+00 2.8668e-02 \\ 
1.9e+00 2.1453e-02 \\ 
2.1e+00 1.5883e-02 \\ 
2.2e+00 1.1636e-02 \\ 
2.4e+00 8.4363e-03 \\ 
2.5e+00 6.0544e-03 \\ 
2.7e+00 4.3016e-03 \\ 
2.8e+00 3.0262e-03 \\ 
3.0e+00 2.1080e-03 \\ 
            };
        \addlegendentry{$\deltaOptLIP_G(\epsilon)$ of Gaussian mechanism~\eqref{eq:Gaussian_mechanism}};

        \addplot [color=red,line width=1.2pt,forget plot]
        table[row sep=crcr]{%
            2.2e-16 5.8848e-02 \\ 
1.6e-01 2.2126e-02 \\ 
3.2e-01 1.3696e-02 \\ 
4.7e-01 1.0550e-02 \\ 
6.3e-01 9.0072e-03 \\ 
7.9e-01 8.1458e-03 \\ 
9.5e-01 7.6337e-03 \\ 
1.1e+00 7.3223e-03 \\ 
1.3e+00 7.1341e-03 \\ 
1.4e+00 7.0236e-03 \\ 
1.6e+00 6.9622e-03 \\ 
1.7e+00 0.0000e+00 \\ 
1.9e+00 6.9160e-03 \\ 
2.1e+00 6.9102e-03 \\ 
2.2e+00 0.0000e+00 \\ 
2.4e+00 0.0000e+00 \\ 
2.5e+00 6.9075e-03 \\ 
2.7e+00 6.9075e-03 \\ 
2.8e+00 0.0000e+00 \\ 
3.0e+00 6.9074e-03 \\ 
            };

        \addplot [color=black,dashed,line width=1.2pt,forget plot]
        table[row sep=crcr]{%
            2.2e-16 0.006907504562964 \\ 
            5 0.006907504562964 \\ 
            };

        \addplot [color=OliveGreen,dashdotted,line width=1.2pt,forget plot]
        table[row sep=crcr]{%
            2.2e-16 1.0419e-01 \\ 
1.6e-01 4.8838e-02 \\ 
3.2e-01 1.8764e-02 \\ 
4.7e-01 5.8644e-03 \\ 
6.3e-01 1.4893e-03 \\ 
7.9e-01 3.0823e-04 \\ 
9.5e-01 5.2276e-05 \\ 
1.1e+00 7.3127e-06 \\ 
1.3e+00 8.4970e-07 \\ 
1.4e+00 8.2659e-08 \\ 
1.6e+00 1.2819e-15 \\ 
1.7e+00 1.2511e-15 \\ 
1.9e+00 1.2019e-15 \\ 
2.1e+00 1.1404e-15 \\ 
2.2e+00 1.1686e-15 \\ 
2.4e+00 1.1642e-15 \\ 
2.5e+00 1.2782e-15 \\ 
2.7e+00 1.0915e-15 \\ 
2.8e+00 1.2428e-15 \\ 
3.0e+00 1.2382e-15 \\ 
            };

            
            
        \coordinate (Center) at (axis cs:.8,1.3e-2);
   		\coordinate (Radius) at (axis cs:.85,3e-2);
   		
   		\coordinate (Center2) at (axis cs:.3,2e-3);
   		\coordinate (Radius2) at (axis cs:.35,5e-3);

        \draw[-latex] (axis cs:1.4,2e-1) node[right] {$\mu = 0.1$} -- (axis cs:1.15,2e-1);
        
        \draw[-latex] (axis cs:.26,8e-4) node[below] {$\mu = 0.01$} -- (axis cs:.45,2.2e-3);
        
    \end{axis}
    \pgfextractx{\radx}{\pgfpointdiff{\pgfpointanchor{Radius}{center}}{\pgfpointanchor{Center}{center}}}%
	\pgfextracty{\rady}{\pgfpointdiff{\pgfpointanchor{Radius}{center}}{\pgfpointanchor{Center}{center}}}%
	\draw (Center) ellipse [x radius = \radx, y radius = \rady]; 
	
	\pgfextractx{\radx}{\pgfpointdiff{\pgfpointanchor{Radius2}{center}}{\pgfpointanchor{Center2}{center}}}%
	\pgfextracty{\rady}{\pgfpointdiff{\pgfpointanchor{Radius2}{center}}{\pgfpointanchor{Center2}{center}}}%
	\draw (Center2) ellipse [x radius = \radx, y radius = \rady]; 
\end{tikzpicture}%
\vspace{-1cm}


    \caption{The achievable \gls{LIP} curve $\deltaLIP_\mu(\epsilon)$ given in~\eqref{eq:delta_mu_LIP} of a $\mu$-\gls{CD}-\revise{\gls{LMIP}} mechanism, its lower bound $1-2^{-\mu}$, and the optimal \gls{LIP} curve $\deltaOptLIP_G(\epsilon)$ of the Gaussian mechanism $G$ in~\eqref{eq:Gaussian_mechanism} with $d = 1$, $\Delta = 1$, $P_X$ being the Gaussian distribution $\normal(0,1/4)$ truncated in a set of $21$ points placed evenly in $[-\Delta,\Delta]$, and $\sigma^2$ chosen such that $\muOptCD_G = \mu$.}
    \label{fig:LMIDP_LIP}
    \vspace{-.3cm}
\end{figure}
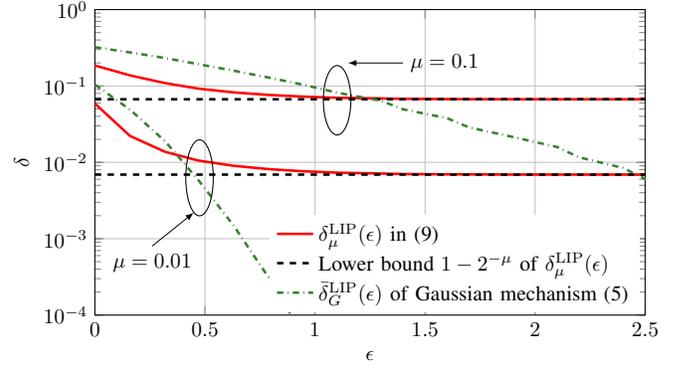

In Fig.~\ref{fig:LMIDP_LIP}, we plot $\deltaLIP_\mu(\epsilon)$ for $\mu \in \{0.1,0.01\}$ and the optimal \gls{LIP} curve of the Gaussian mechanism that also achieves $\mu$-\gls{CD}-\revise{\gls{LMIP}} for a given discrete input distribution. Similar to $\deltaLDP_\mu(\epsilon)$, the general \gls{LIP} guarantee $\deltaLIP_\mu(\epsilon)$ saturates to a lower bound given by $1 - 2^{-\mu}$. On the contrary, the optimal \gls{LIP} curve of the Gaussian mechanism keeps decreasing with~$\epsilon$. This shows that \gls{CD}-\revise{\gls{LMIP}} is a weak privacy notion and does not necessarily imply a strong \gls{LIP} guarantee. 

In Fig.~\ref{fig:LIP_LMIDP}, we show $\muCD_{\delta(\epsilon)}$ for $\delta(\epsilon) = \deltaOptLIP_G(\epsilon)$ and the optimal \gls{CD}-\revise{\gls{LMIP}} parameter $\muOptCD_G$ of the Gaussian mechanism considered in Fig.~\ref{fig:LIP_LMIDP}. We see that $\muCD_{\delta(\epsilon)}$ keeps decreasing as $\delta(\epsilon)$ gets lower. Therefore, a strong \gls{LIP} mechanism also performs well in terms of~\gls{CD}-\revise{\gls{LMIP}}.
\begin{figure}[t]
    \centering
    \begin{tikzpicture}[scale = .8]
    \begin{axis}[%
        width=3.6in,
        height=2in,
        at={(0.759in,0.481in)},
        scale only axis,
        xmin=0,
        xmax=200,
        xlabel style={font=\color{black},yshift=.5ex},
        xlabel={noise variance $\sigma^2$},
        ymin=1e-4,
        ymax=1,
        yminorticks=true,
        ymode=log,
        ytick={1, 1e-1, 1e-2, 1e-3, 1e-4},
        ylabel={$\mu$},
        xmajorgrids,
        ymajorgrids,
        legend style={at={(.99,.99)}, anchor=north east, legend cell align=left, align=left,draw=none, fill=white, fill opacity=.5,text opacity = 1}
        ]		
        
        \addplot [color=red,line width=1.2pt]
        table[row sep=crcr]{%
            1.0e+00 1.9498e+00 \\ 
2.0e+00 5.9591e-01 \\ 
3.0e+00 3.4514e-01 \\ 
4.0e+00 2.4159e-01 \\ 
5.0e+00 1.8533e-01 \\ 
6.0e+00 1.5007e-01 \\ 
7.0e+00 1.2595e-01 \\ 
8.0e+00 1.0843e-01 \\ 
9.0e+00 9.5138e-02 \\ 
1.0e+01 8.4717e-02 \\ 
1.1e+01 7.6320e-02 \\ 
1.2e+01 6.9424e-02 \\ 
1.3e+01 6.3658e-02 \\ 
1.4e+01 5.8766e-02 \\ 
1.5e+01 5.4564e-02 \\ 
1.6e+01 5.0917e-02 \\ 
1.7e+01 4.7722e-02 \\ 
1.8e+01 4.4900e-02 \\ 
1.9e+01 4.2390e-02 \\ 
2.0e+01 4.0143e-02 \\ 
2.1e+01 3.8120e-02 \\ 
2.2e+01 3.6289e-02 \\ 
2.3e+01 3.4625e-02 \\ 
2.4e+01 3.3105e-02 \\ 
2.5e+01 3.1711e-02 \\ 
2.6e+01 3.0471e-02 \\ 
2.7e+01 2.9245e-02 \\ 
2.8e+01 2.8150e-02 \\ 
2.9e+01 2.7132e-02 \\ 
3.0e+01 2.6187e-02 \\ 
3.1e+01 2.5304e-02 \\ 
3.2e+01 2.4477e-02 \\ 
3.3e+01 2.3703e-02 \\ 
3.4e+01 2.2976e-02 \\ 
3.5e+01 2.2291e-02 \\ 
3.6e+01 2.1646e-02 \\ 
3.7e+01 2.1038e-02 \\ 
3.8e+01 2.0463e-02 \\ 
3.9e+01 1.9919e-02 \\ 
4.0e+01 1.9402e-02 \\ 
4.1e+01 1.8911e-02 \\ 
4.2e+01 1.8444e-02 \\ 
4.3e+01 1.7999e-02 \\ 
4.4e+01 1.7574e-02 \\ 
4.5e+01 1.7167e-02 \\ 
4.6e+01 1.6778e-02 \\ 
4.7e+01 1.6408e-02 \\ 
4.8e+01 1.6055e-02 \\ 
4.9e+01 1.5715e-02 \\ 
5.0e+01 1.5390e-02 \\ 
5.1e+01 1.5078e-02 \\ 
5.2e+01 1.4778e-02 \\ 
5.3e+01 1.4490e-02 \\ 
5.4e+01 1.4213e-02 \\ 
5.5e+01 1.3946e-02 \\ 
5.6e+01 1.3689e-02 \\ 
5.7e+01 1.3441e-02 \\ 
5.8e+01 1.3202e-02 \\ 
5.9e+01 1.2971e-02 \\ 
6.0e+01 1.2748e-02 \\ 
6.1e+01 1.2533e-02 \\ 
6.2e+01 1.2352e-02 \\ 
6.3e+01 1.2166e-02 \\ 
6.4e+01 1.1988e-02 \\ 
6.5e+01 1.1739e-02 \\ 
6.6e+01 1.1556e-02 \\ 
6.7e+01 1.1378e-02 \\ 
6.8e+01 1.1206e-02 \\ 
6.9e+01 1.1039e-02 \\ 
7.0e+01 1.0877e-02 \\ 
7.1e+01 1.0719e-02 \\ 
7.2e+01 1.0566e-02 \\ 
7.3e+01 1.0417e-02 \\ 
7.4e+01 1.0273e-02 \\ 
7.5e+01 1.0132e-02 \\ 
7.6e+01 9.9950e-03 \\ 
7.7e+01 9.8617e-03 \\ 
7.8e+01 9.7319e-03 \\ 
7.9e+01 9.6054e-03 \\ 
8.0e+01 9.4821e-03 \\ 
8.1e+01 9.3619e-03 \\ 
8.2e+01 9.2448e-03 \\ 
8.3e+01 9.1305e-03 \\ 
8.4e+01 9.0190e-03 \\ 
8.5e+01 8.9102e-03 \\ 
8.6e+01 8.8039e-03 \\ 
8.7e+01 8.7001e-03 \\ 
8.8e+01 8.5988e-03 \\ 
8.9e+01 8.4998e-03 \\ 
9.0e+01 8.4030e-03 \\ 
9.1e+01 8.3083e-03 \\ 
9.2e+01 8.2158e-03 \\ 
9.3e+01 8.1253e-03 \\ 
9.4e+01 8.0368e-03 \\ 
9.5e+01 7.9501e-03 \\ 
9.6e+01 7.8653e-03 \\ 
9.7e+01 7.7823e-03 \\ 
9.8e+01 7.7010e-03 \\ 
9.9e+01 7.6214e-03 \\ 
1.0e+02 7.5434e-03 \\ 
1.1e+02 6.8425e-03 \\ 
1.2e+02 6.2601e-03 \\ 
1.3e+02 5.7687e-03 \\ 
1.4e+02 5.3484e-03 \\ 
1.5e+02 4.9860e-03 \\ 
1.6e+02 4.6710e-03 \\ 
1.7e+02 4.3926e-03 \\ 
1.8e+02 4.1421e-03 \\ 
1.9e+02 3.9171e-03 \\ 
2.0e+02 3.7176e-03 \\ 
            };
        \addlegendentry{$\muCD_{\delta(\epsilon)}$ in~\eqref{eq:mu_delta_LIP}};

        \addplot [color=OliveGreen,dashdotted,line width=1.2pt]
        table[row sep=crcr]{%
            1.0e+00 1.3285e-01 \\ 
2.0e+00 6.9483e-02 \\ 
3.0e+00 4.7058e-02 \\ 
4.0e+00 3.5578e-02 \\ 
5.0e+00 2.8602e-02 \\ 
6.0e+00 2.3913e-02 \\ 
7.0e+00 2.0545e-02 \\ 
8.0e+00 1.8009e-02 \\ 
9.0e+00 1.6030e-02 \\ 
1.0e+01 1.4443e-02 \\ 
1.1e+01 1.3142e-02 \\ 
1.2e+01 1.2055e-02 \\ 
1.3e+01 1.1135e-02 \\ 
1.4e+01 1.0346e-02 \\ 
1.5e+01 9.6605e-03 \\ 
1.6e+01 9.0605e-03 \\ 
1.7e+01 8.5306e-03 \\ 
1.8e+01 8.0593e-03 \\ 
1.9e+01 7.6374e-03 \\ 
2.0e+01 7.2574e-03 \\ 
2.1e+01 6.9135e-03 \\ 
2.2e+01 6.6007e-03 \\ 
2.3e+01 6.3150e-03 \\ 
2.4e+01 6.0529e-03 \\ 
2.5e+01 5.8118e-03 \\ 
2.6e+01 5.5891e-03 \\ 
2.7e+01 5.3829e-03 \\ 
2.8e+01 5.1913e-03 \\ 
2.9e+01 5.0129e-03 \\ 
3.0e+01 4.8464e-03 \\ 
3.1e+01 4.6906e-03 \\ 
3.2e+01 4.5445e-03 \\ 
3.3e+01 4.4072e-03 \\ 
3.4e+01 4.2779e-03 \\ 
3.5e+01 4.1560e-03 \\ 
3.6e+01 4.0409e-03 \\ 
3.7e+01 3.9320e-03 \\ 
3.8e+01 3.8288e-03 \\ 
3.9e+01 3.7309e-03 \\ 
4.0e+01 3.6378e-03 \\ 
4.1e+01 3.5493e-03 \\ 
4.2e+01 3.4650e-03 \\ 
4.3e+01 3.3846e-03 \\ 
4.4e+01 3.3079e-03 \\ 
4.5e+01 3.2345e-03 \\ 
4.6e+01 3.1644e-03 \\ 
4.7e+01 3.0972e-03 \\ 
4.8e+01 3.0328e-03 \\ 
4.9e+01 2.9710e-03 \\ 
5.0e+01 2.9117e-03 \\ 
5.1e+01 2.8548e-03 \\ 
5.2e+01 2.8000e-03 \\ 
5.3e+01 2.7472e-03 \\ 
5.4e+01 2.6965e-03 \\ 
5.5e+01 2.6475e-03 \\ 
5.6e+01 2.6003e-03 \\ 
5.7e+01 2.5548e-03 \\ 
5.8e+01 2.5108e-03 \\ 
5.9e+01 2.4683e-03 \\ 
6.0e+01 2.4273e-03 \\ 
6.1e+01 2.3875e-03 \\ 
6.2e+01 2.3491e-03 \\ 
6.3e+01 2.3119e-03 \\ 
6.4e+01 2.2758e-03 \\ 
6.5e+01 2.2408e-03 \\ 
6.6e+01 2.2069e-03 \\ 
6.7e+01 2.1741e-03 \\ 
6.8e+01 2.1421e-03 \\ 
6.9e+01 2.1111e-03 \\ 
7.0e+01 2.0810e-03 \\ 
7.1e+01 2.0517e-03 \\ 
7.2e+01 2.0233e-03 \\ 
7.3e+01 1.9956e-03 \\ 
7.4e+01 1.9687e-03 \\ 
7.5e+01 1.9425e-03 \\ 
7.6e+01 1.9169e-03 \\ 
7.7e+01 1.8921e-03 \\ 
7.8e+01 1.8679e-03 \\ 
7.9e+01 1.8442e-03 \\ 
8.0e+01 1.8212e-03 \\ 
8.1e+01 1.7988e-03 \\ 
8.2e+01 1.7769e-03 \\ 
8.3e+01 1.7555e-03 \\ 
8.4e+01 1.7346e-03 \\ 
8.5e+01 1.7142e-03 \\ 
8.6e+01 1.6943e-03 \\ 
8.7e+01 1.6749e-03 \\ 
8.8e+01 1.6558e-03 \\ 
8.9e+01 1.6373e-03 \\ 
9.0e+01 1.6191e-03 \\ 
9.1e+01 1.6013e-03 \\ 
9.2e+01 1.5839e-03 \\ 
9.3e+01 1.5669e-03 \\ 
9.4e+01 1.5503e-03 \\ 
9.5e+01 1.5340e-03 \\ 
9.6e+01 1.5180e-03 \\ 
9.7e+01 1.5024e-03 \\ 
9.8e+01 1.4871e-03 \\ 
9.9e+01 1.4720e-03 \\ 
1.0e+02 1.4573e-03 \\
1.1e+02 1.3250e-03 \\ 
1.2e+02 1.2147e-03 \\ 
1.3e+02 1.1213e-03 \\ 
1.4e+02 1.0413e-03 \\ 
1.5e+02 9.7189e-04 \\ 
1.6e+02 9.1119e-04 \\ 
1.7e+02 8.5762e-04 \\ 
1.8e+02 8.1000e-04 \\ 
1.9e+02 7.6739e-04 \\ 
2.0e+02 7.2904e-04 \\ 
            };
        \addlegendentry{$\muOptCD_G$ of Gaussian mechanism~\eqref{eq:Gaussian_mechanism}};


   		

        
    \end{axis}
	
\end{tikzpicture}%
\vspace{-1cm}
    \caption{The achievable \gls{CD}-\revise{\gls{LMIP}} parameter $\muCD_{\delta(\epsilon)}$ of a mechanism that satisfies $(\epsilon,\delta(\epsilon))$-\gls{LDP} with $\delta(\epsilon) = \deltaOptLIP_G(\epsilon)$, and the optimal \gls{CD}-\revise{\gls{LMIP}} parameter $\muOptCD_G$ 
    of the Gaussian mechanism considered in Fig.~\ref{fig:LMIDP_LIP}.}
    \label{fig:LIP_LMIDP}
    \vspace{-.3cm}
\end{figure}

\section{The Optimality of the Gaussian Mechanism}
While we have shown that \revise{\gls{LMIP}} is a weak privacy notion and thus should not be directly used for mechanism design, \revise{\gls{LMIP}} can still be used as an analysis tool to gain insights into the performance of privacy mechanisms, as done in~\cite{Cuf16}. In this section, we analyze the optimality of the Gaussian mechanism through the lens of \gls{CI}-\revise{\gls{LMIP}}. Specifically, we shall show that for the mechanism
\begin{equation}
    M(\rvecx) = \rvecx + \rvecy 
\end{equation}
with $\rvecx, \rvecy \in \reals^d$, \revise{among all distributions of $\rvecy$ satisfying $\Exp{\|\rvecy\|^2} \le d\sigma^2$, the Gaussian distribution $\rvecy \sim \normal(\mathbf{0}, \sigma^2\matidentity_d)$ (as in~\eqref{eq:Gaussian_mechanism}) achieves the strongest \gls{CI}-\revise{\gls{LMIP}} guarantee.} Here, we consider a relaxed constraint\footnote{This average power constraint is implied by the peak power constraint $\|\rvecx\|^2 \le d\Delta^2$. \revise{(The latter can be} imposed by the norm clipping step typically included in \gls{DP} schemes for machine learning, such as \gls{DP}-SGD~\cite{Aba16}.\revise{)} Therefore, the privacy of a mechanism for input with average power constraint can also be achieved for input with peak power constraint.} $\Exp{\|\rvecx\|^2} \le d\Delta^2$.  Without loss of generality, we further assume that $\rvecx$ and $\rvecy$ have zero mean. 

\noindent The smallest value of $\mu$ such that $M$ satisfies $\mu$-\gls{CI}-\revise{\gls{LMIP}} is
\begin{align}
    &\inf_{P_\rvecy :\; \Exp{\rvecy} = \mathbf{0}, \Exp{\|\rvecy\|^2} \le d\sigma^2} \sup_{P_{\rvecx} :\; \Exp{\rvecx} = \mathbf{0}, \Exp{\|\rvecx\|^2} \le d\Delta^2} \!\!I(\rvecx; \rvecx \!+\! \rvecy) \notag\\
    &= \inf_{\matK_\rvecy \in \setK_d(d\sigma^2)} \inf_{P_\rvecy \in \setP(\matK_\rvecy)} \sup_{\matK_{\rvecx} \in \setK_d(d\Delta^2)} 
    \sup_{P_{\rvecx} \in \setP(\matK_{\rvecx})} \!I(\rvecx; \rvecx \!+\! \rvecy) \label{eq:opt_cov}\\
    &= \inf_{\matK_{\rvecy} \in \setK_d(d\sigma^2)}  \sup_{\matK_{\rvecx} \in \setK_d(d\Delta^2)}  \mu(\matK_{\rvecx},\matK_{\rvecy}) \label{eq:inf_sup}
\end{align}
with 
\begin{equation}
    \mu(\matK_{\rvecx},\matK_\rvecy) = 
    \inf_{P_{\rvecy} \in \setP(\matK_\rvecy)}
    \sup_{P_{\rvecx} \in \setP(\matK_{\rvecx})} I(\rvecx; \rvecx + \rvecy). \label{eq:tmp482}
\end{equation}
Here $\setK_d(a)$ is the space of $d \times d$ symmetric and positive semi-definite matrices with trace upper-bounded by $a$ and $\setP(\matK)$ is the space of vector distributions with mean $\veczero$ and covariance matrix $\matK$. 
Equation~\eqref{eq:inf_sup} requires that $\inf_{P_\rvecy \in \setP(\matK_\rvecy)}$ and $\sup_{\matK_{\rvecx} \in \setK_d(d\Delta^2)}$  in~\eqref{eq:opt_cov} are interchangeable, which we prove Appendix~\ref{proof:inf_sup_interchangeable}.
The solution to~\eqref{eq:inf_sup} is presented next.
\begin{theorem}[Uncorrelated Gaussian is the best-case noise for \gls{CI}-\revise{\gls{LMIP}}] \label{th:best_noise_LMIDP}
    An optimizing pair of distributions $(P_{\rvecx}, P_\rvecy)$ for~\eqref{eq:tmp482} is given by $\big(\normal(\mathbf{0}, \matK_{\rvecx}), \normal(\mathbf{0}, \matK_\rvecy)\big)$, which leads to 
    \begin{align} \label{eq:best_case_noise}
        \mu(\matK_{\rvecx},\matK_\rvecy) = \frac{1}{2} \log\det(\matidentity_d + \matK_\rvecy^{-1} \matK_{\rvecx}).
    \end{align}
    Furthermore, the optimizing covariance matrices in~\eqref{eq:inf_sup} are given by $(\matK_{\rvecx},\matK_\rvecy) = (\Delta^2 \matidentity_d, \sigma^2 \matidentity_d)$.
\end{theorem}
\begin{IEEEproof}
    See Appendix~\ref{proof:best_noise_LMIDP}.
\end{IEEEproof} 


Theorem~\ref{th:best_noise_LMIDP} implies that, for \gls{CI}-\revise{\gls{LMIP}}, under the considered average power constraint, the best-case additive noise follows the uncorrelated Gaussian distribution. 
\revise{This result is related to the capacity of a vector \gls{AWGN} channel. Indeed, it is known that the worst-case additive noise distribution under a covariance constraint is Gaussian~\cite{Dig01}.}

\section{Conclusions}
We provided explicit conversion rules between \revise{\gls{LMIP}} and \gls{LDP} (for the context-independent setting), and between \revise{\gls{LMIP}} and \gls{LIP} (for the context-dependent setting). Our results showed that a strong \revise{\gls{LMIP}} guarantee does not necessarily imply a strong \gls{LDP}/\gls{LIP} guarantee. We therefore highlighted that \revise{\gls{LMIP}} should not be used as a design objective but rather as an analysis tool. We used \revise{\gls{LMIP}} to showcase the optimality of the uncorrelated Gaussian mechanism.

\appendix
We shall use the following result in the proofs. 
\begin{lemma} \label{lemma:KL_hockeystick}
    Given two measures $P$ and $Q$, it holds that 
    \begin{equation}
        \!\KL(P \| Q) = \log(e) \!\int_{0}^\infty\!\! (\opH_{e^\epsilon}(P \| Q) + e^{-\epsilon} \opH_{e^\epsilon}(Q \| P)) \dif \epsilon. \!\label{eq:KL_hockeystick}
    \end{equation}
\end{lemma}
\begin{IEEEproof}
    We define the random variable $L = \ln \frac{P(\rvecy)}{Q(\rvecy)}$ with $\rvecy \sim P$. It follows from~\cite[Thm.~6]{Bal18b} that the moment-generating function of~$L$ can be expressed in terms of the hockey-stick divergence between $P$ and $Q$ as 
    \begin{align}
        &\varphi_L(t) = \notag \\ &1 + t(t + 1) \int_0^\infty \!\! (e^{t \epsilon} \opH_{e^{\epsilon}}(P \| Q) + e^{-(t+1) \epsilon} \opH_{e^{\epsilon}}(Q \| P))  \dif \epsilon. 
    \end{align}
    By definition, 
    $\KL(P \| Q) = \log(e) \Exp[]{L} = \log(e) \varphi'_L(0)$. After some simple manipulations, we obtain that $\varphi'_L(0)$ is given by the integral on the right-hand side of~\eqref{eq:KL_hockeystick}.
\end{IEEEproof}

\subsection{Proof of Theorem~\ref{th:LMIDP_and_LDP}} \label{proof:LMIDP_and_LDP}
    \subsubsection{Part~\ref{th:LMIDP_implies_LDP}}
    Consider a mechanism $M(\rvecx)$ that satisfies $\mu$-\gls{CI}-\revise{\gls{LMIP}}. Consider an arbitrary pair $(\vecx,\vecx')$ and arbitrary event~$\setE$. Denote by $\setP_{\vecx,\vecx'}$ the set of probability distributions in~$\setP$ that only put positive mass on $\vecx$ and~$\vecx'$. For~$\rvecx$ following a distribution in $\setP_{\vecx,\vecx'}$, let $A$ take value~$1$ if $\rvecx = \vecx$ and value~$0$ if $\rvecx = \vecx'$. Furthermore, let $B = \ind{M(\rvecx) \in \setE}$. 
    Since the mechanism satisfies $\mu$-\gls{CI}-\revise{\gls{LMIP}}, we have that
    \begin{align}
        \mu \ge \textstyle\sup_{P_\rvecx \in \setP} I(\rvecx; M(\rvecx)) 
        &\ge \sup_{P_\rvecx \in \setP_{\vecx,\vecx'}} I(\rvecx; M(\rvecx)) \notag\\
        &\ge \textstyle\sup_{P_A} I(A;B), \label{eq:tmp1237}
    \end{align}
    where the last inequality follows by applying the data processing inequality to the Markov chain $A \leftrightarrow \rvecx\leftrightarrow M(\rvecx)\leftrightarrow B$. 
    
    We define 
    \begin{align}
        p_1 &= \Prob{B = 1 \given A = 1} = \Prob{M(\vecx) \in \setE} \\
        p_0 &= \Prob{B = 1 \given A = 0} = \Prob{M(\vecx') \in \setE}.
    \end{align}
    Notice that $\sup_{P_A} I(A;B)$ is the capacity of a binary asymmetric channel with crossover probabilities $(p_0,1-p_1)$. Therefore, $\sup_{P_A} I(A;B)$ is equal to $C\sub{BAC}(p_0,1-p_1)$. 
    From this and~\eqref{eq:tmp1237}, we conclude that $(p_0,p_1)$ satisfies $C\sub{BAC}(p_0,1-p_1) \le \mu$. We define $\deltaLDP_\mu(\epsilon)$ as in~\eqref{eq:delta_mu}. 
    Using the definition of $p_0$ and $p_1$, we deduce that~\eqref{eq:def_LDP} is satisfied with $\delta = \deltaLDP_\mu(\epsilon)$ for every $(\vecx,\vecx')$ and every event $\setE$. Therefore, the mechanism $M$ satisfies $(\epsilon,\deltaLDP_\mu(\epsilon))$-\gls{LDP}. 

    Let $(p_0^*,p_1^*)$ be the maximizer in~\eqref{eq:delta_mu}. For an arbitrary set $\setA$, consider the randomized mechanism $M^*(\rvecx)$ whose output is drawn from the ${\rm Ber}(p_1^*)$ distribution if $\rvecx \in \setA$ and from ${\rm Ber}(p_0^*)$ otherwise. We have that 
    \begin{align}
        \max_{P_\rvecx}I(\rvecx; M^*(\rvecx)) &\le \max_{P_\rvecx} I(\ind{\rvecx \in \setA}; M^*(\rvecx)) \\
        &= C\sub{BAC}(p_0^*,1-p_1^*) \\
        &\le \mu.
    \end{align}
    Therefore, the mechanism $M^*$ satisfies $\mu$-\gls{CI}-\revise{\gls{LMIP}}. However, we can also easily verify that, for every $\epsilon > 0$,
    \begin{equation}
        \max_{\vecx,\vecx'} \max_{\setE \subset \{0,1\}} \big( \Prob{M(\vecx) \in \setE} - e^\epsilon \Prob{M(\vecx') \in \setE} \big) 
        = \deltaLDP_\mu(\epsilon). 
    \end{equation}
    Therefore, the mechanism does not satisfy $(\epsilon,\delta)$-\gls{LDP} for $\epsilon > 0$ and any $\delta < \deltaLDP_\mu(\epsilon)$. 

\subsubsection{Part~\ref{th:LDP_implies_LMIDP}}
    By using Lemma~\ref{lemma:KL_hockeystick}, we have that
    \begin{align}
        \KL(P_{M(\vecx)} \| P_{M(\vecx')})
        &= \log(e)\int_0^\infty \big(\opH_{e^\epsilon}(P_{M(\vecx)} \| P_{M(\vecx')}) \notag \\
        &\quad  + e^{-\epsilon} \opH_{e^\epsilon}(P_{M(\vecx')} \| P_{M(\vecx)})\big) \dif \epsilon \label{eq:tmp484} \\
        &\le \log(e) \int_0^\infty (1 + e^{-\epsilon}) \delta(\epsilon) \dif \epsilon, \label{eq:tmp458}
    \end{align}
    where~\eqref{eq:tmp458} follows from~\eqref{eq:LDP_hockeystick}.
    We now bound the mutual information $I(\rvecx;M(\rvecx))$ for input distribution $P_\rvecx$ as
    \begin{align}
        \!\!\!I(\rvecx; M(\rvecx)) \!&= \Exp[\rvecx\!\!]{\KL(P_{M(\rvecx) \given \rvecx} \; \|\; P_{M(\rvecx)})} \\
        &= \Exp[\rvecx\!\!]{\KL(P_{M(\rvecx) \given \rvecx} \| \Exp[\rvecx'\!\!]{P_{M(\rvecx') \given \rvecx'}})}\!\! \label{eq:tmp492}\\
        &\le \Exp[\rvecx,\rvecx'\!\!]{\KL(P_{M(\rvecx) \given \rvecx} \| P_{M(\rvecx') \given \rvecx'})} \label{eq:tmp493} \\
        &\le \log(e) \int_0^\infty (1 + e^{-\epsilon}) \delta(\epsilon) \dif \epsilon. \label{eq:tmp494}
    \end{align}
    In~\eqref{eq:tmp492}, $\rvecx' \sim P_\rvecx$; \eqref{eq:tmp493} follows from Jensen's inequality and the fact that the \gls{KL} divergence is convex in the second argument; and~\eqref{eq:tmp494} holds because~\eqref{eq:tmp458} holds for every realizations $\vecx$ and $\vecx'$ of $\rvecx$ and $\rvecx'$, respectively. Since~\eqref{eq:tmp494} holds for every input distribution $P_\rvecx$, $M$ satisfies $\muCI_{\delta(\epsilon)}$-\revise{\gls{LMIP}} with~$\muCI_{\delta(\epsilon)}$ given by the right-hand side of~\eqref{eq:tmp494}.

\subsection{Proof of Theorem~\ref{th:LMIDP_and_LIP}} 
\label{proof:LMIDP_and_LIP}

    \subsubsection{Part (a)}
    Since the mechanism satisfies $\mu$-\gls{CD}-\revise{\gls{LMIP}}, we have that, for the input distribution $P_\rvecx$,
    \begin{equation}
        \!\!\mu  \ge I(\rvecx; M(\rvecx)) = \Exp[\rvecx \sim P_\rvecx\!\!]{\KL(P_{M(\rvecx) \given \rvecx} \| P_{M(\rvecx)})}\!.\!\! \label{eq:tmp491}
    \end{equation}
    
    Consider an arbitrary input $\vecx$ and arbitrary event $\setE$. We let $B = \ind{M(\rvecx) \in \setE}$. It follows that $B$ follows the Bernoulli distribution with parameter $p_0 = \Prob{M(\rvecx) \in \setE}$. Furthermore, given $\rvecx = \vecx$,  $B$ follows the Bernoulli distribution with parameter $p_1 = \Prob{M(\vecx) \in \setE}$. Applying the data processing inequality for \gls{KL} divergence~\cite[Thm.~2.15]{PolWu24}, we have that
    \begin{align}
        \KL(P_{M(\rvecx) \given \rvecx = \vecx} \| P_{M(\rvecx)}) 
        &\ge \KL(P_{B \given \rvecx = \vecx} \| P_{B}) \\
        &= \KL(\Ber{p_1} \| \Ber{p_0}). \label{eq:tmp500} 
    \end{align}
    From~\eqref{eq:tmp491} and~\eqref{eq:tmp500}, we obtain that $\KL(\Ber{p_1} \| \Ber{p_0}) \le \mu$. 
    We next define $\deltaLIP_\mu(\epsilon)$ as in~\eqref{eq:delta_mu_LIP}.
    Using the definition of $p_0$ and $p_1$, we get that~\eqref{eq:def_LIP} is achieved with $\delta = \deltaLIP_\mu(\epsilon)$ for every $\vecx$ and every event $\setE$. Therefore, $M$ satisfies $(\epsilon,\deltaLIP_\mu(\epsilon))$-\gls{LIP}. 

    
    
    \subsubsection{Part (b)}
    We follow similar steps as for the proof of~Theorem~\ref{th:LMIDP_and_LDP}\ref{th:LDP_implies_LMIDP}. First, for a given $P_\rvecx$, we have that
    \begin{align}
        \KL(P_{M(\vecx)} \| P_{M(\rvecx)}) 
        &= \log(e)\int_0^\infty \big(\opH_{e^\epsilon}(P_{M(\vecx)} \| P_{M(\rvecx)}) \notag \\
        &\quad  + e^{-\epsilon} \opH_{e^\epsilon}(P_{M(\rvecx)} \| P_{M(\vecx)})\big) \dif \epsilon  \\
        &\le \log(e) \int_0^\infty (e^{\epsilon} + e^{-\epsilon}) \delta(\epsilon) \dif \epsilon, \label{eq:tmp400}
    \end{align}
    where~\eqref{eq:tmp400} follows from~\eqref{eq:LIP_hockeystick}.
    Then, we bound the mutual information $I(\rvecx;M(\rvecx))$ as 
    \begin{align}
        I(\rvecx; M(\rvecx)) &= \Exp[\rvecx]{\KL(P_{M(\rvecx) \given \rvecx} \| P_{M(\rvecx)})} \\
        &\le \log(e) \int_0^\infty (e^{\epsilon} + e^{-\epsilon}) \delta(\epsilon) \dif \epsilon \label{eq:tmp405},
    \end{align}
    where~\eqref{eq:tmp405} holds because~\eqref{eq:tmp400} holds for every realization $\vecx$ of $\rvecx$. 
    It follows that the mechanism $M$ satisfies $\muCD_{\delta(\epsilon)}$-\gls{CD}-\revise{\gls{LMIP}} with $\muCD_{\delta(\epsilon)}$ given by the right-hand side of~\eqref{eq:tmp405}.

\subsection{Proof of~\eqref{eq:inf_sup}} \label{proof:inf_sup_interchangeable}
    According to the minimax theorem~\cite[Thm.~2]{Sim95}, the $\inf_{P_\rvecy \in \setP(\matK_\rvecy)}$ and $\sup_{\matK_\rvecx \in \setK_d(d\Delta^2)}$ in~\eqref{eq:opt_cov} are interchangeable if the following conditions are simultaneously satisfied:
    \begin{enumerate}[label=(\roman*)]
        \item the sets $\setP(\matK_\rvecy)$ and $\setK_d(d\Delta^2)$ are compact and convex, \label{cond:i}
        
        \item the function $f(\matK_\rvecx, P_\rvecy) = \sup_{P_\rvecx \in \setP(\matK_\rvecx)} I(\rvecx; \rvecx + \rvecy)$ is convex over $P_\rvecy \in \setP(\matK_\rvecy)$ for a given $\matK_\rvecx$, \label{cond:ii}
        
        \item $f(\matK_\rvecx, P_\rvecy)$ is concave over $\matK_\rvecx \in \setK_d(d\sigma^2)$ for a given~$P_\rvecy$. \label{cond:iii}
    \end{enumerate}
     Condition~\ref{cond:i} readily holds. Condition~\ref{cond:ii} holds because the mutual information $I(\rvecx; \rvecx + \rvecy)$ is convex in $P_\rvecy$ and because the point-wise supremum preserves convexity~\cite[Sec.~3.2.3]{Boy04}. It remains to verify Condition~\ref{cond:iii}, which we do next.
     
     Let us fix $P_\rvecy$ and consider two covariance matrices $\matK_0$ and $\matK_1$ in $\setK_d(d\Delta^2)$. Let $$P_i = \argmax_{P_\rvecx \in \setP(\matK_i)} I(\rvecx; \rvecx + \rvecy)$$ for $i\in \{0,1\}$. Let $\rvecx_0 \sim P_0$ and $\rvecx_1 \sim P_1$. Furthermore, let $$\rvecx = (1-\theta) \rvecx_0 + \theta \rvecx_1$$ with $\theta \sim \Ber{\alpha}$. 
    We have the expansions
    \begin{align}
        I(\rvecx; \rvecx + \rvecy , \theta) &= I(\rvecx + \rvecy; \theta) + I(\rvecx + \rvecy; \rvecx \given \theta) \\
        &= I(\rvecx + \rvecy; \rvecx) + I(\rvecx + \rvecy; \theta \given \rvecx). \label{eq:tmp1045}
    \end{align}
    Since $\vecx + \rvecy$ is independent of $\theta$ for every $\vecx$, we have that $I(\rvecx + \rvecy; \theta \given \rvecx) = 0$. It then follows from~\eqref{eq:tmp1045} that 
    \begin{equation}
        I(\rvecx + \rvecy; \rvecx \given \theta) \le I(\rvecx + \rvecy; \rvecx). \label{eq:tmp1049}
    \end{equation}
    We expand $I(\rvecx + \rvecy; \rvecx \given \theta)$ as
    \begin{align}
        &I(\rvecx + \rvecy; \rvecx \given \theta) \notag \\ 
        &= (1-\alpha)I(\rvecx_0 + \rvecy; \rvecx_0) + \alpha I(\rvecx_1 + \rvecy; \rvecx_1)  \\
        &= (1-\alpha) f(\matK_0, P_\rvecy) + \alpha f(\matK_1,P_\rvecy), \label{eq:tmp1054}
    \end{align}
    where~\eqref{eq:tmp1054} holds due to the definition of $P_0$ and $P_1$.
    Moreover, since $\rvecx$ has covariance matrix $(1-\alpha)\matK_0 + \alpha\matK_1$, we have
    \begin{align}
        I(\rvecx + \rvecy; \rvecx) 
        &\le f((1-\alpha)\matK_0 + \alpha\matK_1, P_\rvecy). \label{eq:tmp1058}
    \end{align}
    From~\eqref{eq:tmp1049}, \eqref{eq:tmp1054}, and~\eqref{eq:tmp1058}, we conclude that $$(1-\alpha) f(\matK_0, P_\rvecy) + \alpha f(\matK_1,P_\rvecy) \le f((1-\alpha)\matK_0 + \alpha\matK_1, P_\rvecy)$$ for every $\alpha \in [0,1]$. Therefore, Condition~\ref{cond:iii} indeed holds. 

\subsection{Proof of Theorem~\ref{th:best_noise_LMIDP}} \label{proof:best_noise_LMIDP}
    The proof is built upon~\cite{Dig01} and~\cite{Hug88}.
    The optimizing measures $(P_\rvecx^*,P_\rvecy^*)$ for~\eqref{eq:tmp482} are given by a saddle point for $I(\rvecx; \rvecx + \rvecy)$, i.e., they satisfy 
    \begin{equation}
        I(\rvecx; \rvecx + \rvecy^*) \le I(\rvecx^*; \rvecx^* + \rvecy^*) \le I(\rvecx^*; \rvecx^* + \rvecy)
    \end{equation}
    for $\rvecx^* \sim P_\rvecx^*$ and $\rvecy^* \sim P_\rvecy^*$, and for every $P_\rvecx \in \setP(\matK_\rvecx)$ and $P_\rvecy \in \setP(\matK_\rvecy)$. According to~\cite[Thm.~II.1]{Dig01}, there exists a saddle point $(P_\rvecx^*,P_\rvecy^*)$ for $I(\rvecx;\rvecx + \rvecy)$ since this mutual information is concave in $P_\rvecx$ and convex in $P_\rvecy$, and the constraint sets $\setP(\matK_\rvecx)$ and $\setP(\matK_\rvecy)$ are convex. Let $P_{\rvecx_G}^*$ and $P_{\rvecy_G}^*$ be Gaussian distributions with zero mean and the same covariance matrices as $P_{\rvecx}^*$ and $P_{\rvecy}^*$, 
    i.e., $(P_{\rvecx_G}^*,P_{\rvecy_G}^*) = \big(\normal(\mathbf{0}, \matK_\rvecx), \normal(\mathbf{0}, \matK_\rvecy)\big)$. Leveraging the fact that the Gaussian distribution maximizes the entropy for a given covariance, \cite[Thm.~II.1]{Dig01} showed that $(P_{\rvecx_G}^*,P_{\rvecy_G}^*)$ is also a saddle point. With $\rvecx^* \sim P_{\rvecx_G}^*$ and $\rvecy^* \sim P_{\rvecy_G}^*$, it is straightforward that $I(\rvecx^*; \rvecx^* + \rvecy^*)$ is given by the right-hand side of~\eqref{eq:best_case_noise}.

    To solve~\eqref{eq:inf_sup} with $\mu_M(\matK_\rvecx,\matK_
   \rvecy)$ given in~\eqref{eq:best_case_noise}, we follow the footsteps of the proof of~\cite[Thm.~3]{Hug88}. Let $\veclambda_{\rvecx}$ and $\veclambda_{\rvecy}$ be vectors containing eigenvalues of $\matK_\rvecx$ and $\matK_\rvecy$, respectively, sorted in decreasing order. Using~\eqref{eq:best_case_noise} and the eigendecomposition of $\matK_\rvecx$ and $\matK_\rvecy$, we get that
    $$\mu_M(\matK_\rvecx,\matK_\rvecy) = \mu_M(\veclambda_{\rvecx},\veclambda_{\rvecy}) = \frac{1}{2} \sum_{i=1}^d \log\big(1 + \frac{\lambda_{\rvecx,i}}{\lambda_{\rvecy,i}}\big).$$ Then~\eqref{eq:inf_sup} 
    becomes
    \begin{equation}
        \!\!\inf_{\veclambda_{\rvecy} \in \reals^d_+ : \sum_{i=1}^d \lambda_{\rvecy,i} \le d\sigma^2} \sup_{\veclambda_{\rvecx} \in \reals^d_+ : \sum_{i=1}^d \lambda_{\rvecx,i} \le d\Delta^2} \!\!\!\!\!\mu_M(\veclambda_{\rvecx},\veclambda_{\rvecy}). \label{eq:opt_cov_lambda}
    \end{equation}
    Notice that it is without loss of optimality to restrict the constraint sets for $\veclambda_\rvecx$ and $\veclambda_\rvecy$ in~\eqref{eq:opt_cov_lambda} to $$\setS_\rvecx = \big\{\veclambda_\rvecx \in \reals_+^d : \textstyle\sum_{i=1}^d \lambda_{\rvecx,i} = d \Delta^2\big\}$$ and $$\setS_\rvecy = \big\{\veclambda_\rvecy \in \reals_+^d : \textstyle\sum_{i=1}^d \lambda_{\rvecy,i} = d \sigma^2\big\},$$ respectively.  
    A solution $(\veclambda_\rvecx^*, \veclambda_\rvecy^*)$ to~\eqref{eq:opt_cov_lambda} must be a saddle point for the function $\mu_M(\veclambda_{\rvecx},\veclambda_{\rvecy})$, i.e, 
    \begin{align}
        \sup_{\veclambda_{\rvecx} \in \setS_\rvecx} \mu_M(\veclambda_{\rvecx},\veclambda_{\rvecy}^*) &= \mu_M(\veclambda_{\rvecx}^*,\veclambda_{\rvecy}^*), 
        \label{eq:sp_lambda_x} \\
        \inf_{\veclambda_{\rvecy} \in \setS_\rvecy} \mu_M(\veclambda_{\rvecx}^*,\veclambda_{\rvecy}) &= \mu_M(\veclambda_{\rvecx}^*,\veclambda_{\rvecy}^*). 
        \label{eq:sp_lambda_z}
    \end{align}

    Using the Gallager's conditions~\cite[Thm.~4.4.1]{Gal68}, we verify that $\veclambda_\rvecx^* = \Delta^2 \vecone_d$ and $\veclambda_\rvecx^* = \sigma^2 \vecone_d$ satisfy~\eqref{eq:sp_lambda_x} and~\eqref{eq:sp_lambda_z}.
    We conclude that $(\Delta^2 \vecone_d, \sigma^2 \vecone_d)$ is the solution to~\eqref{eq:opt_cov_lambda}, i.e., $(\matK_\rvecx,\matK_\rvecy) = (\Delta^2 \matidentity_d, \sigma^2 \matidentity_d)$ is the solution to~\eqref{eq:opt_cov}.
\bibliographystyle{IEEEtran}
\bibliography{IEEEabrv,./biblio}

\end{document}